\documentclass{ws-ijmpd}
\usepackage[super,compress]{cite}
\begin{document}

\markboth{James M. Nester and Chiang-Mei Chen}
{Gravity: a gauge theory perspective}

%
\catchline{}{}{}{}{}
%

\title{Gravity: a gauge theory perspective}

\author{James M. Nester$^{1,2,3,4,5}$ and Chiang-Mei Chen$^{1,2,6}$}

\address{$^1$Department of Physics, National Central University, Chungli 32001, Taiwan\\
$^2$Center for Mathematical and Theoretical Physics, National Central University\\ Chungli 32001, Taiwan\\
$^3$Graduate Institute of Astronomy, National Central University,\\ Chungli 32001, Taiwan\\
$^4$Leung Center for Cosmology and Particle Astrophysics,\\
National Taiwan University, Taipei 10617, Taiwan\\
$^5$nester@phy.ncu.edu.tw\\
$^6$cmchen@phy.ncu.edu.tw}

\maketitle

\begin{history}
\received{Day Month Year}
\revised{Day Month Year}
\end{history}

\begin{abstract}
The evolution of a generally covariant theory is under-determined.  One hundred years ago such  dynamics had never before been considered; its ramifications were perplexing,  its future important role for all the fundamental interactions under the name  gauge principle could not be foreseen.  We recount some history regarding Einstein, Hilbert, Klein and Noether and the novel features of gravitational energy that led to Noether's two theorems.  Under-determined evolution is best revealed in the Hamiltonian formulation.  We developed a covariant Hamiltonian formulation. The Hamiltonian boundary term gives covariant expressions for the quasi-local energy, momentum and angular momentum.  Gravity can be considered as a gauge theory of the local Poincar\'e group.  The dynamical potentials of the Poincar\'e gauge theory of gravity  are the frame and the connection.  The spacetime geometry has in general both curvature and torsion.  Torsion naturally couples to spin;  it could have a significant magnitude and yet not be noticed, except on a cosmological scale where it could have significant effects.
\end{abstract}


\ccode{PACS numbers: 04.20.Cv, 04.20.Fy}

\section{Introduction}
There have long been disputes about all of the \emph{principles\/} used by Einstein for his gravity theory.
 Kretschmann in 1917 argued that \emph{general covariance\/} has no real physical content and no connection to an extension of the principle of relativity.\cite{Norton93,Norton03}
From a different perspective general covariance has deep fundamental ramifications.

GR with general covariance is the premier gauge theory. The consequences of this,
especially regarding  gravitational energy and under-determined evolution, were long perplexing.    The Hamiltonian approach clarifies these issues.  Gravity can be understood as a gauge theory of the local Poincar\'e symmetries of spacetime.

\section{Some Historical Background}

In 1915 Einstein made  presentations to the Prussian Academy of Sciences on Nov.~4, 11, 18 and 25 (published one week later). The last had his generally covariant equations along with energy conservation.
Hilbert presented his ``{Foundations of physics}'' on November 16 and 20;
he submitted his first note on Nov.~19 (published in March 1916).\cite{Sauer99}
His first theorem is central to our concerns:

\smallskip
\noindent
\textbf{Theorem I} ({`Leitmotiv'}\footnote{guiding theme}).
\textit{In the system of $n$ Euler-Lagrange differential equations in $n$ variables obtained from a generally covariant variational integral such as in Axiom I, 4 of the $n$ equations are always a consequence of the other $n - 4$ in the sense that 4 linearly independent combinations of the $n$ equations and their total derivatives are always identically satisfied.}

Some discussions of Hilbert's work have appeared recently,\cite{Sauer99,Rowe99,Corry04} often it has been viewed from the Einstein GR perspective (e.g., Ref.~\refcite{RS07}).
An alternative  view\cite{BR08} of Hilbert's agenda
 argues that a main aim was to reconcile the tension between \emph{general covariance\/} and its inevitable consequence: \emph{a lack of unique determinism\/}\footnote{Einstein had struggled with this in connection with his hole argument.}---this is
\emph{the essence of gauge theory}.
Dynamical equations obtained from a variational principle had formerly had deterministic Cauchy initial value problems, but for GR there was a differential identity connecting the evolution equations; they were not independent and could not give uniquely determined evolution.
Later it was found that this is best addressed using the Hamiltonian approach.\cite{Dirac64,Earman03}

We note some excerpts from Einstein's correspondence, from  Vol.~8 in Ref.~\refcite{CPAE}.

``In your paper everything is understandable to me now except for the energy theorem. Please do not be angry with me that I ask you about this again. \dots
How is this cleared up? It would suffice, of course, if you would charge Miss Noether with explaining this to me.''
 (Doc.~223 to Hilbert 30 May 1916)

 ``The only thing I was unable to grasp in your paper is the conclusion at the top of page 8 that $\varepsilon^\sigma$ was a vector.''
 (Doc.~638 to Klein 22 Oct 1918)
 ``\dots Meanwhile, with Miss Noether's help, I understand that the proof for the vector character of $\varepsilon^\sigma$ {from ``higher principles'' as I had sought was already given by Hilbert on pp.\ 6, 7 of his first note}, \dots'' (Doc.~650 from Klein 10 Nov 1918)

Briefly, after a couple of years Klein clarified Hilbert's energy-momentum ``vector''; he related it to Einstein's pseudotensor, but he disagreed with Einstein's physical interpretation of divergenceless expressions.\footnote{Things were not as easy then; in particular the Bianchi identity and its contracted version were not known to these people;\cite{Pais82,Rowe02} they each had to rediscover an equivalent identity on their own.}
Enlisted by Hilbert and Klein, it was Emmy Noether who resolved the primary puzzle regarding gravitational energy.

\section{Automatic Conservation of the Source and Gauge Fields}

 In 1916 Einstein showed that local coordinate invariance plus his field equations gives material energy momentum conservation, without using the matter field equations (see Doc.~41 in Vol.~6 of Ref.~\refcite{CPAE}).
This is referred to as \emph{automatic conservation of the source\/} (see section 17.1 in Ref.~\refcite{MTW});  it uses  a {Noether second theorem} local (gauge) symmetry type of argument to obtain current conservation.
Hermann Weyl argued in this way for the electromagnetic current in his  papers of 1918 (the name  \emph{gauge theory\/} comes from this work) and 1929,\footnote{An English translation of Weyl's seminal papers can be found in Ref.~\refcite{Dawn}.} whereas modern field theory generally uses Noether's first theorem for current conservation.\cite{Brad02}.

The \emph{essence} of gauge theory is \emph{a local symmetry}, consequently:
(i) a differential identity,
(ii)  under-determined evolution,
(iii) restricted type of source coupling,
(iv) automatic conservation of the source.
{Yang-Mills is only one special type}.
Our gauge approach to gravity does not try to force it into the Yang-Mills mold, but rather simply recognizes the natural symmetries of spacetime geometry.

\section{Noether's 1918 Contribution}

One word well describes 20th century physics: \emph{symmetry}.
Most of the theoretical physics ideas involved symmetry---essentially they are applications of Noether's two theorems.\cite{KS-Noether}
The first associates conserved quantities with global symmetries.
The second concerns local symmetries: it is the foundation of the modern gauge theories.

Why did Noether make her investigation?
Klein was looking into the relationship between Einstein's pseudotensor and Hilbert's energy vector. He published a paper based on his correspondence with Hilbert.  We quote some excerpts:\cite{KS-Noether}

Klein: ``You know that Miss Noether advises me continually regarding my work, and that in fact it
is only thanks to her that I have understood these questions.''

Hilbert: ``I fully agree in fact with your statements on the energy theorems: Emmy Noether, on whom
I have called for assistance more than a year ago to clarify this type of analytical questions
concerning my energy theorem, found at that time that the energy components that I
had proposed---as well as those of Einstein---could be formally transformed, \dots
into expressions whose divergence
vanishes identically,\dots
Indeed I believe that in the case of general relativity, i.e., in the case of the general invariance
of the Hamiltonian function, the energy equations which in your opinion correspond
to the energy equations of the theory of orthogonal invariance do not exist at all; I can even
call this fact a characteristic of the general theory of relativity.''

This is why Noether wrote her paper.
After presenting her two famous theorems she uses them to draw the conclusion that clarifies the situation:\cite{KS-Noether}

 ``Given $I$ invariant under the group
of translations, then the energy relations are improper if and only if $I$ is invariant
under an infinite group which contains the group of translations as a subgroup.
\dots
As Hilbert expresses his assertion, the lack of a proper law of energy constitutes
a characteristic of the ``general theory of relativity.'' For that assertion to be literally
valid, it is necessary to understand the term ``general relativity'' in a wider sense than
is usual, and to extend it to the aforementioned groups that depend on $n$ arbitrary
functions.''

Her result regarding the lack of a proper law of energy applies not just to Einstein's GR, but to all geometric theories of gravity.
The modern view is that energy-momentum is \emph{quasi-local}, associated with a closed 2 surface.\cite{Sza09}.

\section{Energy-momentum Pseudotensors and the Hamiltonian}

The Einstein Lagrangian differs from Hilbert's by a total divergence:
\begin{eqnarray}
2\kappa{\cal L}_{\rm E}(g_{\alpha\beta}, \partial_\mu g_{\alpha\beta}):=
-\sqrt{-g}g^{\beta\sigma} \Gamma^\alpha{}_{\gamma\mu} \Gamma^\gamma{}_{\beta\nu}\delta^{\mu\nu}_{\alpha\sigma}
\equiv
\sqrt{-g}R-\hbox{div}.
\end{eqnarray}
The Einstein \emph{pseudotensor\/} is the associated canonical energy-momentum density:
\begin{equation}\label{EpseudoT}
\mathfrak{t}_{\rm E}^\mu{}_\nu := \delta^\mu_\nu{\cal L}_{\rm E} - \frac{\partial{\cal L}_{\rm E}}{\partial \partial_\mu g_{\alpha\beta}} \partial_\nu g_{\alpha\beta}.
\end{equation}
Using $\sqrt{-g}G^\mu{}_\nu=\kappa\mathfrak{T}^\mu{}_\nu$ one gets a
 conserved total energy-momentum:
\begin{equation}
\partial_\mu (\mathfrak{T}^\mu{}_\nu+\mathfrak{t}_{\rm E}^\mu{}_\nu)=0,
\qquad\Longleftrightarrow\qquad
\sqrt{-g}G^\mu{}_\nu+\kappa\mathfrak{t}_{\rm E}^\mu{}_\nu=\partial_\lambda \mathfrak{U}^{[\mu\lambda]}{}_\nu.
\end{equation}
The superpotential was found by Freud in 1939:\cite{Freud}
$\mathfrak{U}_{\rm F}^{\mu\lambda}{}_\nu:=-\mathfrak{g}^{\beta\sigma}
\Gamma^\alpha{}_{\beta\gamma}\delta^{\mu\lambda\gamma}_{\alpha\sigma\nu}
$.
Other pseudotensors
 likewise follow from different superpotentials.  They are all inherently reference frame dependent.
Thus there are two big problems: (1) which pseudotensor? (2) which reference frame?
The Hamiltonian approach has answers.

With constant $Z^\mu$, the energy-momentum within a region is
\begin{eqnarray}
-Z^\mu P_\mu(V) &:=& - \int_V Z^\mu ({{\mathfrak{T}}^\nu{}_\mu+{\mathfrak{t}}^\nu{}_\mu}) \sqrt{-g}d^3\Sigma_\nu
\nonumber\\
&\equiv& \int_V \left[ Z^\mu \sqrt{-g} \left( \frac1{\kappa} G^\nu{}_\mu - T^\nu{}_\mu \right) - \frac1{2\kappa} \partial_\lambda \left( Z^\mu {\mathfrak{U}^{\nu\lambda}{}_\mu} \right) \right] d^3\Sigma_\nu
\nonumber\\
&\equiv& \int_V Z^\mu {\cal H}^{\rm GR}_\mu + \oint_{S=\partial V} {\cal B}^{\rm GR}(Z) \equiv H(Z, V). \label{basicHam}
\end{eqnarray}
${\cal H}^{\rm GR}_\mu$ is the covariant expression for the {Hamiltonian density}.
  The {\em boundary term\/} 2-surface integral
 is determined by the superpotential.
The value of the
pseudotensor/Hamiltonian is \emph{quasi-local}, from just the boundary term, since by the initial value constraints the spatial volume integral vanishes.

\section{The Hamiltonian Approach}

Noether's work
  can be combined with the Hamiltonian formulation.
In Hamiltonian field theory, the conserved currents are the generators of the associated symmetry.
For spacetime translations (infinitesimal diffeomorphisms), the associated  current expression (i.e., the energy-momentum density) \emph{is\/} the Hamiltonian density---the canonical generator of spacetime displacements.
   Because it can be varied one gets a handle on the conserved current ambiguity.   The Hamiltonian variation gives information that tames the ambiguity in the boundary term---namely boundary conditions.  Pseudotensor values are values of the Hamiltonian with certain boundary conditions.\cite{CNC99} Thus Problem (1) is under control.

The Hamiltonian approach reveals certain aspects of a theory.
The constrained Hamiltonian formalism was developed by Dirac\cite{Dirac64} and by Bergmann and coworkers.  It was applied to GR by Pirani, Schild and Skinner\cite{PSS} and by Dirac\cite{Dirac58}.  Later the ADM approach\cite{ADM} became dominant.
For the Poincar\'e gauge theory of gravity (PG) the Hamiltonian approach was developed by Blagojevi\'c and coworkers.\cite{BN83}

\section{The Covariant Hamiltonian and its Boundary Term}
From a first order Lagrangian formulation, ${\cal L}=d\varphi\wedge p-\Lambda$, which gives pairs of first order equations for an $f$-form $\varphi$ and its conjugate $p$,  we developed a 4D-\emph{covariant\/} Hamiltonian formalism.\cite{CNT95,CN99,CNC99,CN00,CNT05,GR100}
The Hamiltonian generates the evolution of a spatial region along a vector field.
The Hamiltonian density is the first order translational Noether current 3-form,
 it is linear in the displacement vector plus a total differential:
\begin{equation}
{\cal H}(Z) := \pounds_Z \varphi \wedge p - i_Z {\cal L} =: Z^\mu {\cal H}_\mu + d {\cal B}(Z),
 \label{5:HN}
\end{equation}
and is a conserved ``current'' {\it on shell\/} (i.e., when the field equations are satisfied):
\begin{equation} \label{5:IdH}
- d {\cal H}(Z) \equiv \pounds_Z \varphi \wedge \frac{\delta {\cal L}}{\delta \varphi} + \frac{\delta {\cal L}}{\delta p} \wedge \pounds_Z p.
\end{equation}
 Furthermore, from {\em local\/} diffeomorphism invariance, it follows that ${\cal H}_\mu$ is linear in the Euler-Lagrange expressions. Hence the translational Noether current conservation reduces to a differential identity.
This an instance of Noether's 2nd theorem, exactly the case to which Hilbert's ``lack of a proper energy law'' remark applies. The value of the Hamiltonian is \emph{quasi-local\/} (associated with a closed 2-surface):
\begin{equation}
-P(Z, V) =H(Z,V) := \int_V {\cal H}(Z) = \oint_{\partial V} {\cal B}(Z). \label{5:EN}
\end{equation}
The Hamiltonian boundary term has two important roles:
(i) it gives the quasi-local values,
(ii) it gives the boundary conditions.
 The boundary term can be adjusted to match suitable boundary conditions.
 We were led to a set of general boundary terms which are linear in $\Delta\varphi:=\varphi-\bar\varphi$, $\Delta p:=p-\bar p$, where $\bar\varphi,\bar p$ are reference values:
\begin{equation}
{\cal B}(Z) := i_Z \left\{ \begin{array}{c} {\varphi} \\ {\bar\varphi} \end{array} \right\} \wedge \Delta p - (-1)^f \Delta\varphi \wedge i_Z \left\{ \begin{array}{c} p \\ \bar p \end{array} \right\}.\label{genB}
\end{equation}
The associated variational Hamiltonian boundary term is
\begin{equation}
\delta{\cal H}(Z) \sim d\left[ \left\{ i_Z \delta\varphi \wedge \Delta p \atop - i_Z \Delta\varphi \wedge \delta p \right\} + (-1)^f \left\{ -\Delta\varphi \wedge i_Z \delta p \atop \delta\varphi \wedge i_Z \Delta p \right\} \right]. \label{deltaHZbound}
\end{equation}
Here \emph{for each bracket independently\/} one may choose either the upper or lower term, which represent essentially a choice of Dirichlet (fixed field) or Neumann (fixed momentum) boundary conditions for the space and time parts of the fields separately.\footnote{There are more complicated possibilities, ``mixed'' choices involving some linear combination of the upper and lower expressions.}

For asymptotically flat spaces the Hamiltonian is \emph{well defined\/}, i.e., the boundary term in its variation vanishes and the quasi-local quantities are well defined at least on the phase space of fields satisfying Regge-Teitelboim like asymptotic conditions:
\begin{equation}
\Delta \varphi \approx {\cal O}^+(1/r) + {\cal O}^-(1/r^2), \qquad \Delta p \approx {\cal O}^-(1/r^2) + {\cal O}^+(1/r^3). \label{5:asymptotics}
\end{equation}
Also the formalism has natural boundary term related energy flux expressions.\cite{CNT05}

\section{Gauge and Geometry}

For the history of gauge theory see Ref.~\refcite{Dawn}.
Gravity as a gauge theory was pioneered by Utiyama (1956, 1959), Sciama (1961) and Kibble (1961).
For accounts of gravity as a spacetime symmetry gauge theory, see Hehl and coworkers\cite{HHKN,Hehl80,HHMN95,GFHF96}, Mielke\cite{MieE87} and Blagojevi\'c\cite{Blag02}.  A comprehensive reader with  summaries, discussions and reprints has recently appeared.\cite{BlagHehl}
 For the observational constraints on torsion see Ni.\cite{Ni10}

GR can be seen as the original gauge theory:  the first physical theory where a local gauge freedom (general covariance) played a key role.
Although the electrodynamics potentials with their gauge freedom were known long before GR
 yet this gauge invariance was not seen as having any important role in connection with the nature of the interaction, the conservation of current,  or a differential identity---until the seminal work of Weyl, which post-dated (and was inspired by) GR.

 We also note the developments of the concept of a  connection in geometry by
   Levi-Civita, Weyl, Schouten, Cartan, Eddington, and others.
{Riemann-Cartan geometry} (with a metric and a metric compatible connection, having both curvature and torsion) is the most appropriate for a dynamic spacetime geometry theory: {its local symmetries are just those of the local Poincar\'e group}.
The conserved quantities, energy-momentum and angular momentum/center-of-mass momentum are associated with  the  Minkowski spacetime symmetry, i.e., the Poincar\'e group.

\section{Riemann-Cartan Geometry and PG Dynamics}
It is natural to consider gravity as a gauge theory of the local Poincar\'e group.  The spacetime geometry that suits this perspective is Riemann-Cartan geometry, which has a (Lorentz signature) metric and a \emph{metric compatible connection\/}: $Dg_{\mu\nu}\equiv0$.

The translation and Lorentz gauge potentials are, respectively,
the \emph{coframe\/} $\vartheta^\alpha=e^\alpha{}_k dx^k$ and \emph{connection\/} $\Gamma^\alpha{}_\beta=\Gamma^\alpha{}_{\beta k}dx^k$ one-forms.
The associated field strengths are the \emph{torsion\/} and \emph{curvature\/} 2-forms:
\begin{eqnarray}
T^\alpha := D \vartheta^\alpha := d \vartheta^\alpha + \Gamma^\alpha{}_\beta \wedge \vartheta^\beta &=& \frac12 T^\alpha{}_{ij} dx^i \wedge dx^j, \qquad\\
\label{tor2}
R^\mu{}_\nu := d \Gamma^\mu{}_\nu + \Gamma^\mu{}_\lambda \wedge \Gamma^\lambda{}_\nu&=&\frac12 R^\mu{}_{\nu ij} dx^i \wedge dx^j. \qquad
\label{curv2}
\end{eqnarray}
The first and second
{Bianchi identities} are
$DT^\alpha\equiv R^\alpha{}_\beta\wedge\vartheta^\beta$ and $DR^\alpha{}_\beta\equiv0$.
The {Ricci identity},
$\left[ \nabla_\mu, \nabla_\nu \right] V^\alpha = R^\alpha{}_{\beta\mu\nu} V^\beta - T^\gamma{}_{\mu\nu} \nabla_\gamma V^\alpha$,
reflects the holonomy and the Lorentz and translational field strengths.
For an orthonormal frame $g_{\mu\nu}=\hbox{const}$ and $\Gamma^{\alpha\beta}$ is \emph{antisymmetric}.

The PG dynamics has been discussed in detail in Ref.~\refcite{GR100} including
(i) the Lagrangian, both 2nd and 1st order,
(ii) the Noether symmetries, conserved currents and differential identities,
(iii) the covariant Hamiltonian including the generators of the local Poincar\'e gauge symmetries,
(iv) our {preferred Hamiltonian boundary term},
(v) the quasi-local energy-momentum and angular momentum/center-of-mass moment obtained therefrom, and
(vi) the {choice of reference} in the boundary term.

\section{Preferred Hamiltonian Boundary Terms and Reference}

For the PG and GR our preferred Hamiltonian boundary terms are
\begin{equation}
{\cal B}_{\rm PG}(Z) = i_Z \vartheta^\alpha \tau_\alpha + \Delta \Gamma^\alpha{}_\beta \wedge i_Z \rho_\alpha{}^\beta + {\bar {D}}_\beta Z^\alpha \Delta \rho_\alpha{}^\beta, \label{Bpref(Z)}
\end{equation}
\begin{equation}
{\cal B}_{\rm GR}(Z) = \frac{1}{2\kappa} (\Delta\Gamma^{\alpha}{}_{\beta} \wedge i_Z \eta_{\alpha}{}^{\beta} + \bar
D_{\beta} Z^\alpha \Delta\eta_{\alpha}{}^\beta), \qquad \eta^{\alpha\beta\dots} := * (\vartheta^\alpha \wedge \vartheta^\beta \wedge \cdots). \label{BprefGR}
\end{equation}
Like many other choices, at spatial infinity the latter gives the standard values for
energy-momentum and angular momentum/center-of-mass momentum.

Our preferred GR expression has some special virtues:
(i) at null infinity it gives the Bondi-Trautman energy and the Bondi energy flux, 
(ii) it is covariant,
(iii) it is positive---at least for spherical solutions and large spheres,
(iv) for small spheres it is a positive multiple of the Bel-Robinson tensor,
(v) first law of thermodynamics for black holes,
(vi)  for spherical solutions it has the hoop property,
(vii) for a suitable choice of reference it vanishes for Minkowski space.

Regarding the second ambiguity inherent in our quasi-local energy-momentum expressions: the choice of reference.  Minkowski space is the natural choice, but one needs to choose a specific Minkowski space.  Recently we proposed
(i) \emph{4D isometric matching on the boundary},\footnote{The hardest part of 4D isometric matching is the embedding of the 2D surface $S$ into Minkowski space; Yau and coworkers have extensively investigated this.\cite{WangYau,CWY15}}
and (ii) \emph{energy optimization\/}
as criteria for selecting the ``best matched'' reference on the boundary of the quasi-local region.
 A detailed discussion of our covariant Hamiltonian boundary terms and our reference choice proposal was presented in the MS parallel session\cite{CLNS15} and in Ref.~\refcite{SCLN15}.
 They have been tested on spherically symmetric and axisymmetric spacetimes.\cite{SCLN14}

\section{The Poincar\'e Gauge Theory of Gravity}
The standard PG Lagrangian density has a quadratic field strength form:\footnote{ $\kappa:=8\pi G/c^4$ and $\varrho^{-1}$ has the dimensions of action.
 $\Lambda$ is the cosmological constant.}
\begin{eqnarray}\label{quadraticL}
{\cal L}_{\rm PG} \sim \frac{1}{\kappa}\left(\Lambda+ \text{curvature}
  +\text{torsion}^2\right) + \frac{1}{\varrho}\,\text{curvature}^2\,.
\end{eqnarray}
Varying $\vartheta,\Gamma$ gives quasi-linear 2nd order dynamical equations for the potentials: 
\begin{eqnarray}
\kappa^{-1}(\Lambda+ \hbox{curv} + D \hbox{ tor} + \hbox{tor}^2)+ \varrho^{-1}\hbox{curv}^2&=& \hbox{energy-momentum},\qquad\\
\kappa^{-1}\hbox{tor}+\varrho^{-1} D\hbox{ curv}&=& \hbox{spin}.
\end{eqnarray}
The general theory has 11 scalar plus 7 pseudoscalar parameters, but
there is one even parity and two odd total differentials; effectively
15 ``physical'' parameters.\cite{BHN,BH}

Torsion couples to intrinsic spin, not orbital angular momentum.\cite{HOP13}
 But highly polarized spin density is practically nonexistent in the present day universe.  So on ordinary scales matter hardly excites or responds directly to torsion.  Torsion could have a significant magnitude and yet be hardly observable: ``dark torsion''.

At very high densities it a different story; at around $10^{52}$ gm/cm$^3$ the nucleon spin density is comparable to the material energy density, {and beyond that the spin-torsion interaction dominates gravity in the PG}. So one can expect major effects in the early universe.  But even in the present day,
 while being hardly noticeable on the lab, solar system, or galactic scale, the gravitational effects of torsion (like $\Lambda$) could well have measurable effects on the cosmological scale.

\section{General PG Homogeneous and Isotropic Cosmologies}
The general PG homogeneous and isotropic cosmology has been considered recently.\cite{HCNY15}
For such cosmologies the general PG has an effective Lagrangian.  From this with $\dot a=aH$, 6 first-order equations for $a,H$,  the scalar and pseudoscalar curvatures $R,X$ and the two ``scalar'' torsion components $u,x$ were obtained:
\begin{eqnarray}
-\frac{w_{4+6}}2\dot R-\frac{\mu_{3-2}}4 \dot
X&=&-\left[-3 \tilde a_2-w_{4+6}R-\frac{\mu_{3-2}}2
X\right]u 
+\left[6\tilde\sigma_2-\frac{\mu_{3-2}}2 R+w_{2+3}X\right]x\nonumber\\
&&+w_{4-2}[2X-24(H-u)x]x, \label{dotRdotX'} 
\\
-\frac{\mu_{3-2}}4 \dot
R+\frac{w_{2+3}}2\dot X&=&-\left[6\tilde\sigma_2-\frac{\mu_{3-2}}2 R+w_{2+3}X\right]u 
+\left[12\tilde a_3+w_{4+6}R+\frac{\mu_{3-2}}2
X\right]x\nonumber\\
&&-w_{4-2}(2R-12[(H-u)^2-x^2+k a^{-2}])x, \label{dotXdotR'}\\ 
\dot H -\dot u&=&\frac{R}6-2H^2+3Hu-u^2+x^2-ka^{-2},\label{Hudot'}
\\
a_2\dot u &=&\frac13(
-a_0R-\tilde\sigma_2X+\rho-3p+4\Lambda) 
+a_2(u^2-3Hu)-4a_3x^2, \\
\qquad\dot x&=&\frac{X}6-3Hx+2xu\label{xdot'}.
\label{udotrho3p'}
\end{eqnarray}
Here the
material energy density satisfies a generalized Friedmann relation:
\begin{eqnarray}
\rho
&=&-\Lambda+3a_0[(H-u)^2-x^2+ka^{-2}]\nonumber\\&&-\frac32 a_2(u^2-2Hu)
+6 a_3x^2+6\tilde\sigma_2x(H-u)\nonumber\\
&&\qquad+\frac{w_{4+6}}{24}\left[R^2-12R\left\{(H-u)^2-x^2+ka^{-2}\right\}\right] \nonumber\\
&&\qquad+\frac{\mu_{3-2}}{24}\left[RX-6X\left\{(H-u)^2-x^2+ka^{-2}\right\}-12Rx(H-u)\right]\nonumber\\
&&\qquad
-\frac{w_{2+3}}{24}\left[X^2-24Xx(H-u)\right]\label{density}.
\end{eqnarray}
The above equations are \emph{the most general---they include all the quadratic PG cosmologies}.
Generically the model has, in addition to the usual metric scale factor, effectively two dynamical ``scalar'' torsion components carrying spin 0$^+$ and 0$^{-}$.
  Typically they, and the Hubble expansion rate, show damped oscillations.
 But the model has other types of behavior in the various degenerate special cases, including GR.\cite{HCNY15}

\section*{Acknowledgments}

This was prepared while J.M.N. was visiting the Morningside Center of Mathematics, Chinese Academy of Sciences, Beijing 100190, China.  C.M.C. was supported by the Ministry of Science and Technology of the R.O.C.
under the grant MOST 102-2112-M-008-015-MY3.


\end{document}